\documentclass[preprint]{emulateapj}

\newcommand{\target}{K2-28}
\usepackage{amsmath,amssymb}
\usepackage{graphicx}
\usepackage{natbib}

\usepackage{aas_macros} 
\slugcomment{}
\shortauthors{Hirano et al.}
\shorttitle{ESPRINT III: A Close-in Super-Earth around a Mid-M Dwarf}
\begin{document}
\title{The K2-ESPRINT Project III: A Close-in Super-Earth around a Metal-rich Mid-M Dwarf}
\author{
Teruyuki Hirano\altaffilmark{1}, 
Akihiko Fukui\altaffilmark{2}, 
Andrew W. Mann\altaffilmark{3,4}, 
Roberto Sanchis-Ojeda\altaffilmark{5,6},
Eric Gaidos\altaffilmark{7}, 
Norio Narita\altaffilmark{8,9,10}, 
Fei Dai\altaffilmark{11},
Vincent Van Eylen\altaffilmark{12},
Chien-Hsiu Lee\altaffilmark{13},
Hiroki Onozato\altaffilmark{14},
Tsuguru Ryu\altaffilmark{9},
Nobuhiko Kusakabe\altaffilmark{10},
Ayaka Ito\altaffilmark{15},
Masayuki Kuzuhara\altaffilmark{1,8},
Masahiro Onitsuka\altaffilmark{9},
Misako Tatsuuma\altaffilmark{16},
Grzegorz Nowak\altaffilmark{17,18},
Enric Pall\`{e}\altaffilmark{17,18},
Ignasi Ribas\altaffilmark{19},
Motohide Tamura\altaffilmark{8,10,16},
Liang Yu\altaffilmark{11}
} 
\altaffiltext{1}{Department of Earth and Planetary Sciences, Tokyo Institute of Technology,
2-12-1 Ookayama, Meguro-ku, Tokyo 152-8551, Japan}
\email{hirano@geo.titech.ac.jp}
\altaffiltext{2}{Okayama Astrophysical Observatory, National Astronomical Observatory of Japan, Asakuchi, Okayama 719-0232, Japan}
\altaffiltext{3}{Hubble Fellow}
\altaffiltext{4}{Department of Astronomy, The University of Texas at Austin, Austin, TX 78712, USA}
\altaffiltext{5}{Department of Astronomy, University of California, Berkeley, CA 94720}
\altaffiltext{6}{NASA Sagan Fellow}
\altaffiltext{7}{Department of Geology \& Geophysics, University of Hawai'i at M\={a}noa, Honolulu, HI 96822, USA}
\altaffiltext{8}{National Astronomical Observatory of Japan, 2-21-1 Osawa, Mitaka, Tokyo 181-8588, Japan}
\altaffiltext{9}{SOKENDAI (The Graduate University for Advanced Studies), 2-21-1 Osawa, Mitaka, Tokyo 181-8588, Japan}
\altaffiltext{10}{Astrobiology Center, National Institutes of Natural Sciences, 2-21-1 Osawa, Mitaka, Tokyo 181-8588, Japan}
\altaffiltext{11}{Department of Physics, and Kavli Institute for Astrophysics and Space Research, Massachusetts Institute of Technology, Cambridge, MA 02139}
\altaffiltext{12}{Stellar Astrophysics Centre, Department of Physics and Astronomy, Aarhus University, Ny Munkegade 120, DK-8000 Aarhus C, Denmark}
\altaffiltext{13}{Subaru Telescope, National Astronomical Observatory of Japan, 650 North Aohoku Place, Hilo, HI 96720, USA}
\altaffiltext{14}{Astronomical Institute, Graduate School of Science, Tohoku University, 6-3 Aramaki Aoba, Aoba-ku, Sendai, Miyagi 980-0857, Japan}
\altaffiltext{15}{Graduate School of Science and Engineering, Hosei University, 3-7-2 Kajino-cho, Koganei-shi, Tokyo 184-8584, Japan}
\altaffiltext{16}{Department of Astronomy, Graduate School of Science, The University of Tokyo, Hongo 7-3-1, Bunkyo-ku, Tokyo, 113-0033}
\altaffiltext{17}{Instituto de Astrof\'{i}sica de Canarias (IAC), 38205 La Laguna, Tenerife, Spain}
\altaffiltext{18}{Departamento de Astrof\'{i}sica, Universidad de La Laguna (ULL), 38206 La Laguna, Tenerife, Spain}
\altaffiltext{19}{Institut de Ci\`{e}ncies de l'Espai (CSIC-IEEC), Carrer de Can Magrans, Campus UAB, 08193 Bellaterra, Spain}

\begin{abstract}
We validate a $R_p=2.32\pm 0.24R_\oplus$ planet on a
close-in orbit ($P=2.260455\pm 0.000041$ days) around \target\ (EPIC 206318379), a metal-rich
M4-type dwarf in the Campaign 3 field of the K2 mission. Our follow-up observations 
included multi-band transit observations from the optical
to the near infrared, low-resolution spectroscopy, and high-resolution adaptive-optics (AO) imaging.  
We perform a global fit to all the observed transits using a Gaussian
process-based method and show that the transit depths in all 
passbands adopted for the ground-based transit follow-ups ($r'_2, z_\mathrm{s,2}, J, H, K_\mathrm{s}$) 
are within $\sim 2\sigma$ of the K2 value.  
Based on a model of the background stellar population and the absence of nearby sources in our AO
imaging, we estimate the probability that a background eclipsing
binary could cause a false positive to be $< 2\times 10^{-5}$.  
We also show that \target\ cannot have a physically associated companion 
of stellar type later than M4, based on the measurement of almost identical transit depths 
in multiple passbands. There is a low probability for a M4 dwarf companion ($\approx 0.072_{-0.04}^{+0.02}$), 
but even if this were the case, the size of \target b falls within the planetary regime.
\target b has the same radius (within $1\sigma$) and experiences a similar irradiation 
from its host star as the well-studied GJ~1214b. 
Given the relative brightness of \target\ in the near infrared
($m_\mathrm{Kep}=14.85$ mag and $m_H=11.03$ mag) and relatively deep transit ($0.6-0.7\%$),
a comparison between the atmospheric properties of these two planets
with future observations would be especially interesting.
\end{abstract}

\section{Introduction\label{s:intro}}\label{s:intro}

With relatively low masses and small physical sizes, M dwarfs are
attractive targets for the search and characterization of small
planets.  GJ 1214b is one of the most intensely observed exoplanets,
and the first detailed atmospheric characterization of this
intermediate-sized planet between Earth- and Neptune-like ones
was enabled by the large transit depth and brightness
of its host star \citep[e.g.,][]{2009Natur.462..891C,
  2010Natur.468..669B, 2014Natur.505...69K}.  However, both the census
of small planets around mid-to-late M dwarfs (stars with effective
temperatures of $T_\mathrm{eff}\leq 3400$ K) and the atmospheric
characterization of these objects are still in their infancy; only
five transiting systems (GJ 1214, GJ 1132, Kepler-42, Kepler-445, and 
Kepler-446) have been reported to date around mid-to-late M dwarfs
\citep{2012ApJ...747..144M, 2015ApJ...801...18M, 2015Natur.527..204B}, and the latter three
are too faint or the transit depths too shallow to permit intensive
follow-up studies.

The failure of a second reaction wheel ended the {\it Kepler} prime
mission, but the spacecraft's second mission (``K2") consists of
observations of new fields and new targets with a cycle of $\sim 80$
days \citep{2014PASP..126..398H}. K2 has so far unveiled many
planetary systems with distinguishing characteristics, including a
compact multi-planet system with sub-Saturn-mass planets in the 3:2
mean motion resonance \citep{2015A&A...582A..33A}, multiple systems
with Earth- to super-Earth-sized planets around rather bright M dwarfs
\citep{2015ApJ...804...10C, 2015ApJ...811..102P}, and a disintegrating
minor planet around a white dwarf \citep{2015Natur.526..546V}.  
In order to fully exploit the worldwide ground facilities for K2 follow-ups, 
we had started the
ESPRINT ({\it Equipo de Seguimiento de Planetas Rocosos Intepretando sus Transitos})
collaboration, which aims to discover and characterize unique transiting
planets unveiled by K2. 
ESPRINT has contributed to the K2 haul of discoveries
with a disintegrating ultra-short period planet with a cometary tail
around an M dwarf \citep[ESPRINT I:][]{2015ApJ...812..112S}, and
confirmation of three systems in the Campaign 1 field by radial
velocity measurements \citep[ESPRINT II:][]{2016arXiv160201851V}.

In this paper we validate a $R_p=2.32R_E$ planet around a mid-M star in K2
Campaign 3, which is located at a high galactic latitude ($+60^\circ$)
and samples a distinct stellar population from the prime {\it Kepler} mission.  
Our target is labeled as EPIC 206318379 (which we call \target\ hereafter) 
with the Kepler magnitude of $m_\mathrm{Kep}=14.854$. 
The relative brightness of the star in the near infrared (e.g., $m_H=11.03$) summarized
in Table \ref{hyo1} from the SDSS \citep{2012ApJS..203...21A} and 2MASS
\citep{2006AJ....131.1163S} catalogs suggest that it is a rather cool
star ($T_\mathrm{eff}\sim 3000$ K) and thus an excellent target for
further follow-up studies.

We organize the rest of the paper as follows. 
In Section \ref{s:obs}, we describe how we reduced and detected the planet candidates in 
K2 field 3, and present the ground-based observations for the target, 
including follow-up transit observations with the Infrared Survey Facility (IRSF) 1.4-m telescope
and Okayama 1.88-m telescope, low-resolution spectroscopy with UH88
and the NASA Infrared Telescope Facility (IRTF), 
and adaptive-optics (AO) imaging with the Subaru 8.2-m telescope. 
We then carefully analyze the ground and space-based transit observations 
in Section \ref{s:analysis} using Gaussian processes. 
The transit depths from the ground-based follow-ups are compared to that from the K2
reduced light curve, and consequently we show that they are all in agreement
within $\sim 2\sigma$. 
Section \ref{s:astrovalid} describes to what extent we are able to exclude the false positive scenario 
by first computing the probability that the transit-like signal is caused by 
a background eclipsing binary, which turns out to be very low ($\approx 2\times 10^{-5}$). 
The almost constant transit depths from the optical to the near infrared also 
put a constraint on the magnitude of possible close-in companion around \target. 
Finally, Section \ref{s:discussion} is devoted to discussion and summary.

\section{Observations and Data Reductions\label{s:obs}}\label{s:obs}
\subsection{K2 Photometry \label{s:K2}}\label{s:K2}
The images of all the K2 Campaign 3 targets were downloaded from MAST, and we used our own tools 
\citep[described in detail in][]{2015ApJ...812..112S} 
to produce corrected light curves ready for our planet search routines. In particular, field 3 targets 
were observed for 69 days, from 2014 November 15 through 2015 January 23. The dataset was split 
into 9 segments with a duration of approximately 7.5 days each. 
Because of the faintness of \target, we defined an aperture at each segment with all the pixels that had 4\% more counts than the mean background on at least 50\% of the images of that segment. The light curves are corrected both for instrumental noise induced by the motion of the telescope and astrophysical long term variability using independent 4th-order polynomials with both time and centroid motion as variables. 

We searched the processed long cadence light curves in K2 Campaign 3 for 
planet candidates using a Box-fitting Least Squares routine 
\citep[BLS:][]{2002A&A...391..369K, 2010ApJ...713L..87J}.
We improved the efficiency of the original BLS by implementing the optimal frequency sampling described in 
\citet{2014A&A...561A.138O}.  
\target\ emerged as a clear detection with SNR of $\sim$9. 
A linear ephemeris analysis gave a best-fit period of
$P=2.26023\pm 0.00012$ days and mid-transit time of $T_{c,0}=2456977.9924\pm 0.0021$.

\begin{figure*}[t]
\begin{center}
\includegraphics[width=16cm]{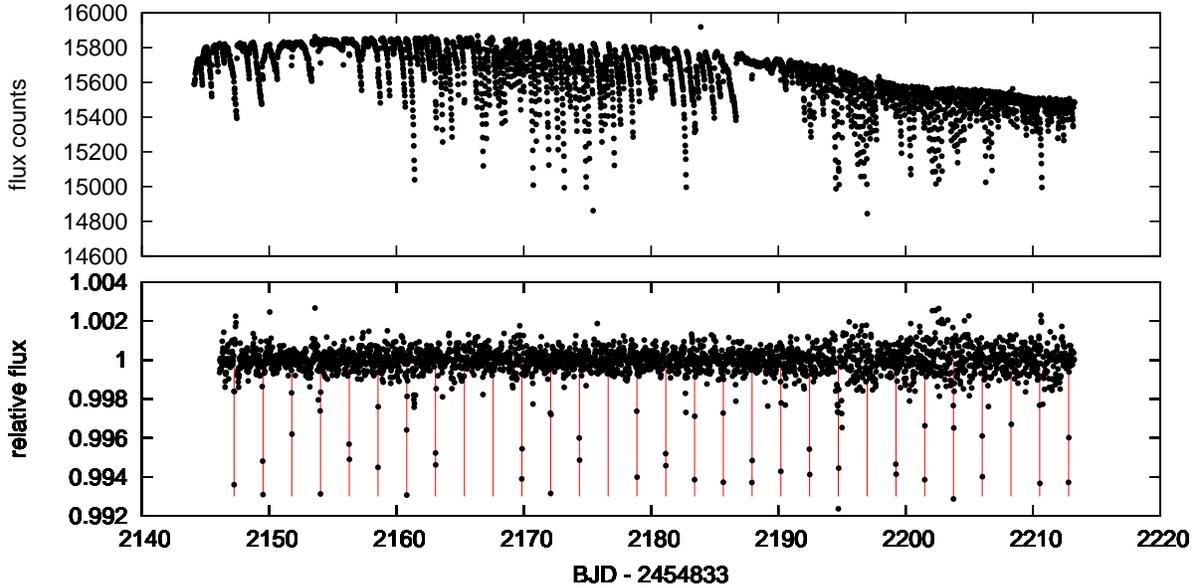} 
\caption{Full light curves for \target\ obtained by our custom-made aperture. 
({\it Top}) Raw light curve. ({\it Bottom}) Reduced light curve after the correction 
for the centroid motion and flux baseline variations. 
The equally-spaced red vertical lines corresponds to the transits of \target b. }
\label{fig:fullcurve}
\end{center}
\end{figure*}
In the archived SDSS image of \target\ there is a second, fainter ($\Delta m_r=4.802\pm 0.035$ mag)
source $\sim 5\farcs2$ to the north-east. 
We made sure that given the difference in magnitude, there is no excess of brightness detected in the K2 data pixels 
around the location of the faint companion seen in the SDSS image. 
Our imaging and astrometric analysis show the fainter source to have proper motion with respect to \target\ and is thus physically unrelated and probably a background star (Section \ref{s:astrometry}). We determined the position of the star at the epoch of K2 observations, designed a K2 photometric aperture that excludes this star but includes the 8 brightest pixels for \target\, and extract a light curve using this aperture and the public code\footnote{https://github.com/vincentvaneylen/k2photometry} outlined in 
\citet{2016arXiv160201851V}.  
Figure \ref{fig:fullcurve} shows thus extracted light curves with and without the correction for
the centroid motions and baseline flux variations. Some of the data points, including ones during transits, 
are missing in the reduced light curve (bottom), mainly due to the removal of outliers 
when we corrected for the centroid motion and baseline function. 
The revised light curve contains the same transit events with the same depth, 
suggesting that \target\ and not the fainter star is the source of the signal.

\begin{figure}[t]
\begin{center}
\includegraphics[width=8.5cm]{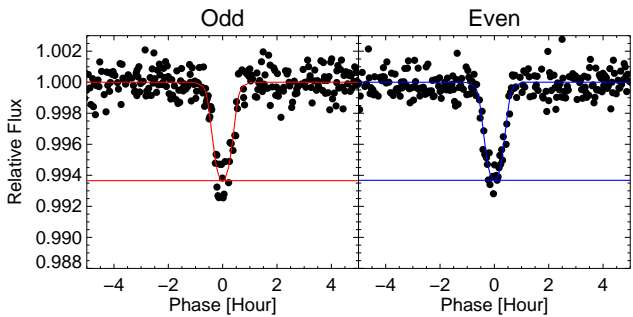} 
\caption{Folded light curves during the odd (left) and even (right) transits. }
\label{fig:oddeven}
\end{center}
\end{figure}

We also performed an odd-even test by folding the K2 light curve with
twice the period of \target\ b. As shown in Figure \ref{fig:oddeven},
the odd and even transits exhibit equal depths within $0.08\sigma$,
where $\sigma$ is defined as the depth uncertainty for each
folded transit in the preliminary depth measurement,
indicating that we have identified the correct period.  The odd-even test also shows no
indication of a secondary eclipse, excluding many false-positive
scenarios involving an eclipsing binary.  We therefore conducted a
campaign of follow-up observations to validate this candidate planet.

\subsection{Follow-up Transit Observations}\label{s:followup}

\subsubsection{IRSF 1.4 m/SIRIUS}
\label{sec:irsf_obs}

We conducted a follow-up transit observation with the Simultaneous Infrared Imager for Unbiased Survey 
\citep[SIRIUS;][]{2003SPIE.4841..459N}
mounted on the IRSF 1.4-m telescope at South African Astronomical Observatory on 2015 August 7 UT. SIRIUS has three infrared detectors, each having 1,024 $\times$ 1,024 pixels with the pixel scale of 0\farcs45 pixel$^{-1}$, allowing $J$-, $H$-, and $K_\mathrm{s}$-band simultaneous imaging. The exposure times were set to 30~s for all bands. We started the observation at 20:50 UT and continued it until 24:03 UT (the airmass had changed from 1.54 to 1.10), covering the expected duration of a transit.
During the SIRIUS observation, we employed the position-locking software
to fix the centroid of stellar images within a few pixels from
the initial position \citep{2013PASJ...65...27N, 2013ApJ...773..144N}.

\begin{figure}[t]
\begin{center}
\includegraphics[width=8.5cm]{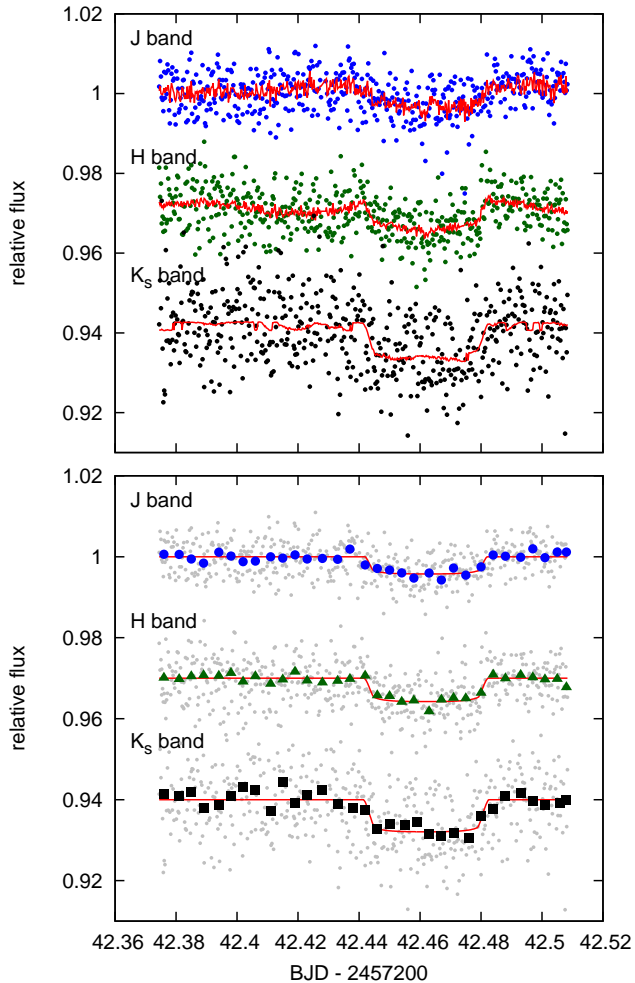} 
\caption{
({\it Top}) \target's raw light curves by IRSF/SIRIUS.
The GP regression to each dataset is shown by the red line. 
({\it Bottom}) Reduced light curves (grey points) after subtracting the GP regressions to
correlated noises. 
The blue, green, and black symbols indicate 16-point binned fluxes
for the $J$, $H$, $K_s$ bands, respectively. 
The best-fit transit model for each band is shown by the red solid line. 
Light curves in different bands have a vertical offset for clarity.}
\label{fig:irsf_jhk}
\end{center}
\end{figure}

The observed images were bias-subtracted and flat-fielded in the
standard manner.  For the flat fielding, we combined 15, 18, and 17
twilight flat images for $J$, $H$, and $K_\mathrm{s}$ bands,
respectively, taken before and after the observation.  We then
performed aperture photometry using two, one, and one comparison
star(s) for $J$, $H$, and $K_\mathrm{s}$, respectively, by using a
custom code \citep{2011PASJ...63..287F}.  
The aperture sizes of 6.0,
8.0, and 6.0 pixels were selected for the $J$-, $H$-, and
$K_\mathrm{s}$-band data to minimize the dispersion of the light
curves with respect to the best-fit transit models. The time system of
the light curves was converted from Julian Day (JD) to Barycentric JD (BJD) 
by the code of \citet{2010PASP..122..935E}. 
The resulting light curves are plotted in the top panel of Figure \ref{fig:irsf_jhk}.

\subsubsection{OAO 188 cm/MuSCAT}

We also conducted a follow-up transit observation with the Multicolor
Simultaneous Camera for studying Atmospheres of Transiting exoplanets
\citep[MuSCAT;][]{2015JATIS...1d5001N} mounted on the 188-cm telescope
at Okayama Astrophysical Observatory (OAO) on 2015 August 23
UT. MuSCAT consists of three CCDs, each having 1,024 $\times$ 1,024
pixels with the pixel scale of 0\farcs36 pixel$^{-1}$, allowing
simultaneous three-band imaging through the Generation 2 Sloan
$g'_2$-, $r'_2$-, and $z_\mathrm{s,2}$-band 
filters\footnote{http://www.astrodon.com/sloan.html}. 
MuSCAT can also fix the centroid of stellar images within $\sim1$ pixel from the
initial position \citep{2015JATIS...1d5001N}.  Because the $g'_2$-band
channel was not available at that time due to an instrumental trouble,
we used the remaining two channels for the observation. The exposure
times were set to 120~s and 60~s for the $r'_2$ and $z_\mathrm{s,2}$
bands, respectively. We started the observation at 17:10 UT and
continued it until 18:48 UT (the airmass had changed from 1.54 to
2.33), covering the first half of the transit.

\begin{figure}[t]
\begin{center}
\includegraphics[width=8.5cm]{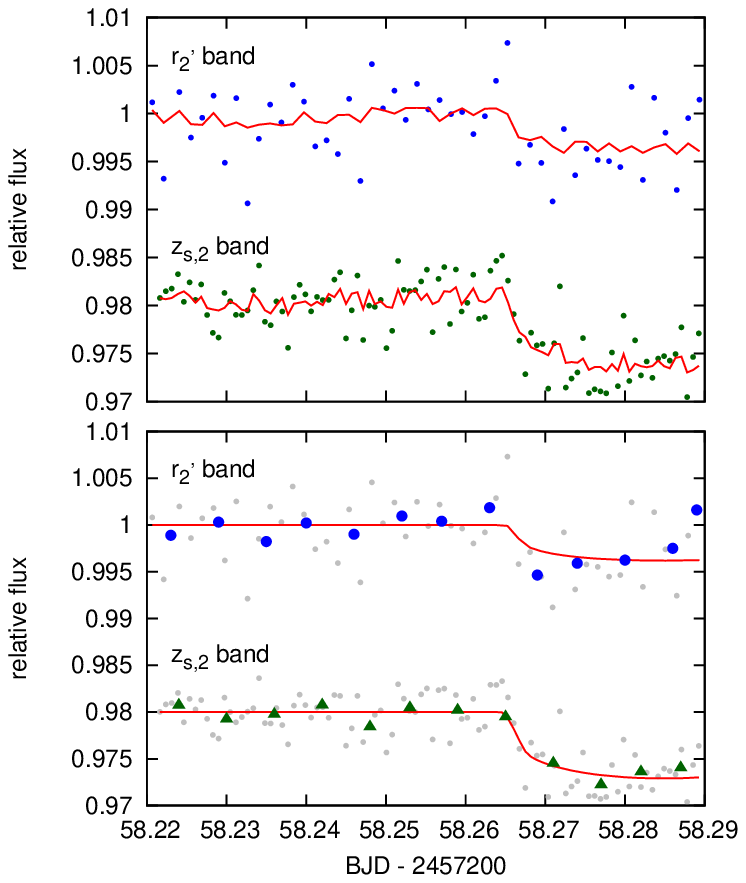} 
\caption{
({\it Top}) \target's raw light curves by OAO/MuSCAT.
The GP regression to each dataset is shown by the red line. 
({\it Bottom}) Reduced light curves (grey points) after subtracting the GP regressions to
correlated noises. 
The blue and green symbols indicate 4-point and 8-point binned fluxes
for the $r'_2$ and $z_\mathrm{s,2}$ bands, respectively. 
The best-fit transit model for each band is shown by the red solid line. 
Light curves in different bands have a vertical offset for clarity.}
\label{fig:oao_rz}
\end{center}
\end{figure}

The observed images were reduced by the same procedure as done in
Section \ref{sec:irsf_obs}.  For the flat fields, we used one hundred
dome-flat images taken on the same observing night for all bands.  We
performed the aperture photometry using three comparison stars for all
bands, with the aperture sizes of 18 pixels for both of the $r'_2$-
and $z_\mathrm{s,2}$-band data. 
The produced light curves are shown in the top panel of 
Figure \ref{fig:oao_rz}.

We note that the comparison stars of our ground-based photometry are 
all solar-type ones with their colors ranging
from $V-J=1.2$ to $2.4$, which are in stark contrast to the color of 
\target\ ($V-J=4.37$). This difference could be a source of systematic effects
in the reduced light curves arising from relative flux variations by e.g., 
changing precipitable water vapor and target's airmass. 
But the five photometric bands that we employed are generally less affected by 
the telluric water absorption, and in particular, the $z_\mathrm{s,2}$-band is
designed to avoid strong telluric absorption. 
The target's airmass was changing monotonically during the IRSF and OAO runs, 
and thus its impact is not expected to be large as long as the baseline of the light curve
is corrected from the out-of-transit flux data.

\subsection{Optical Low Resolution Spectroscopy\label{s:spectroscopy}}\label{s:spectroscopy}

We obtained an optical spectrum of \target\ with the SuperNova Integral Field Spectrograph \citep[SNIFS][]{Lantz2004} on the UH88 telescope on Mauna Kea during the night of August 9, 2015 (UT).  SNIFS provides spectra covering 3200 to 9700 \AA\ at a resolution of $\lambda/\Delta\lambda=700-1000$. Full details on data reduction can be found in \citet{Aldering2002} and \citet{Mann2013}. The resulting spectrophotometric calibration of SNIFS spectra has been shown to be good to 2-3\% \citep{Mann2013}, which 
is sufficient to establish fundamental stellar parameters 
of M dwarfs.

\begin{table}[tb]
\begin{center}
\caption{Stellar Parameters of \target\ (EPIC 206318379)}\label{hyo1}
\begin{tabular}{lc}
\hline
Parameter & Value \\\hline\hline
\multicolumn{2}{l}{\it (Stellar Parameters from the SDSS and 2MASS Catalogs)} \\
RA & $22:22:29.88$ \\
Dec & $-07:57:19.55$\\
$\mu_\alpha$ (mas yr$^{-1}$) & $-256\pm3$  \\
$\mu_\delta$ (mas yr$^{-1}$) & $-195\pm 3$\\
$m_g$ (mag) & $16.839\pm 0.004$ \\
$m_r$ (mag) & $15.449\pm 0.004$ \\
$m_i$ (mag) & $13.909\pm 0.003$ \\
$m_z$ (mag) & $13.103\pm 0.003$ \\
$m_J$ (mag) & $11.70\pm 0.03$ \\
$m_H$ (mag) & $11.03\pm 0.02$ \\
$m_{K_\mathrm{s}}$ (mag) & $10.75\pm 0.02$ \\
\hline
\multicolumn{2}{l}{\it (Spectroscopic and Derived Parameters)} \\
$T_{\rm eff}$ (K) & $3214\pm 60$ \\
$[\mathrm{Fe/H}]$ (dex) & $0.26\pm0.10$ \\
$\log g$ (dex) & $4.93\pm0.04$\\
$R_\star$ ($R_\odot$) & $0.288\pm 0.028$\\
$M_\star$ ($M_\odot$) & $0.257\pm0.048$ \\
$\rho_\star$ ($\rho_\odot$) & $10.8\pm 2.4$ \\
distance (pc) & $53\pm 8$\\
\hline
\end{tabular}
\end{center}
\end{table}

We derived the effective temperature following the procedure from
\citet{Mann2013}. To briefly summarize, we compare our optical
spectrum to BT-SETTL models \citep{Allard2013} calibrated to reproduce
the temperatures of nearby stars with radii and temperatures measured
from long-baseline interferometry \citep{Boyajian2012}. 
This method yielded $T_{\rm{eff}}=3214\pm60$ K. 

\subsection{Near Infrared Spectroscopy\label{s:irtf}}\label{s:irtf}
We obtained a near-infrared (NIR) spectrum of \target\ with the updated SpeX spectrograph \citep{Rayner2003} mounted on IRTF on Mauna Kea. SpeX observations were taken with the 0.3$\times15\arcsec$ slit in the cross-dispersed mode, which provides simultaneous coverage from 0.8 to 2.4$\mu$m at $R\simeq2000$. We placed the target at two positions along the slit (A and B) and observed in an ABBA pattern in order to subsequently subtract the sky background. In total we took six exposures following this pattern, which when stacked, yielded a S/N per pixel of 65 in the $H$ and $K$ bands. To correct for telluric lines, an A-type star was observed immediately after the target observations with much higher S/N ($>100$). 

SpeX spectra were extracted using the SpeXTool package \citep{Cushing2004}, which includes flat-field correction, wavelength calibration, sky subtraction, and extraction of the one-dimensional spectrum. Multiple exposures were combined using the IDL code \textit{xcombspec}.  A telluric correction spectrum was constructed from the A0V star and applied using the \textit{xtellcor} package \citep{Vacca2003}. 

Metallicity was calculated from the IRTF spectrum utilizing procedure from \citet{Mann2013a}. \citet{Mann2013a} provides empirical relations between spectroscopic feature strength and metallicity calibrated against wide binaries containing a solar-type star and an M dwarf. We used the mean of the $H$ and $K$ band relations, accounting for both Poisson and calibration errors. This method gave a metallicity of $[\mathrm{Fe/H}]=0.26\pm0.10$.
With $T_{\rm{eff}}$ derived from the optical spectrum and $[\mathrm{Fe/H}]$ from the NIR spectrum, 
we computed the stellar radius and mass following the empirical relations from \citet{Mann2015b}.  
These values are reported in Table \ref{hyo1} 
along with other derived parameters 
(i.e., the stellar density $\rho_\star$, surface gravity $\log g$, and distance to the star).

\subsection{AO Imaging\label{s:ircs}}\label{s:ircs}

We conducted a high-angular resolution imaging with the Subaru
telescope equipped with the adaptive optics (AO) instrument AO188 and
the Infrared Camera and Spectrograph (IRCS, Kobayashi et al. 2000) on
2015 September 17 UT. We used the ``high-resolution'' mode of IRCS,
which has a pixel scale of 20.6 mas pixel$^{-1}$ and the FOV of
21\farcs1 $\times$ 21\farcs1. We used the target star itself as a
natural guide star. The target star was observed through the $K'$-band
filter at nine dithering points, each with the exposure time of 30~s
(2~s $\times$ 15 coadds), resulting in the total integration time of
270~s.  The airmass was 1.26 and the AO-worked full width at half
maximum (FWHM) of the target 
$\sim$0\farcs18.

\begin{figure}[t]
\begin{center}
\includegraphics[width=8.5cm]{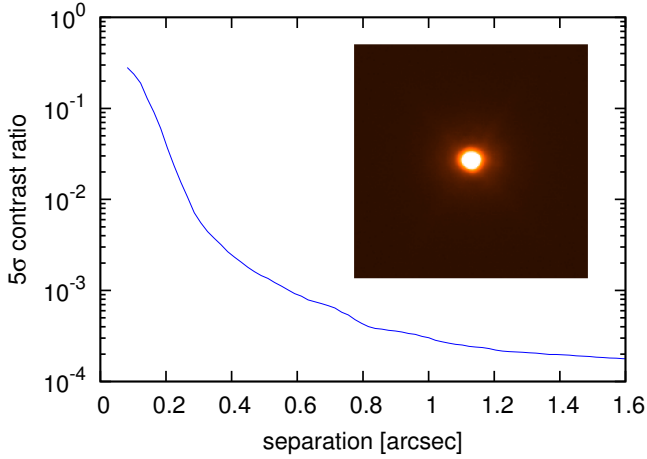} 
\caption{$5\sigma$ Contrast curve around \target\ based on the AO observation with Subaru/IRCS. 
The inset displays the combined image with the FOV of $3\farcs2\times3\farcs2$. North is up and east is left.}
\label{fig:ircs}
\end{center}
\end{figure}

The observed images were dark-subtracted and flat-fielded in a standard manner. 
Twilight flat images taken in the morning were used for the flat fielding. 
The reduced images were then aligned, sky-level-subtracted, and median-combined. 
The combined image and the 5-$\sigma$ contrast curve are shown in Figure \ref{fig:ircs}.

\section{Analyses of the Light Curves\label{s:analysis}}\label{s:analysis}

\begin{figure}[t]
\begin{center}
\includegraphics[width=8.5cm]{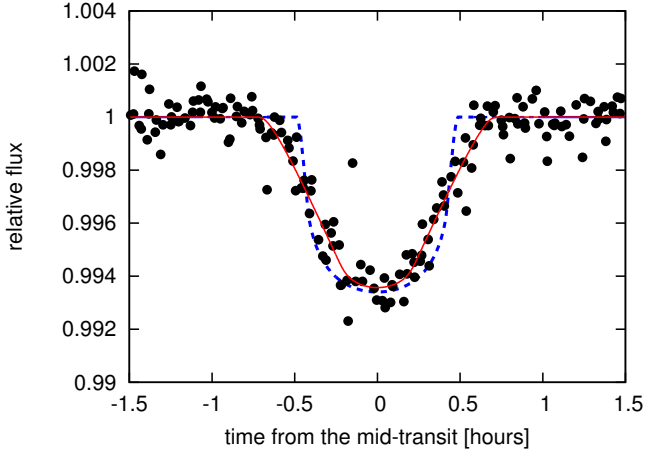} 
\caption{Phase-folded K2 light curve (black points) around the transit of \target b. 
The best-fit transit model, integrated over the cadence of K2 observation ($\sim 29.4$ minutes),
is shown by the red solid line. 
The transit model before the time-integration is shown by the blue dashed line. 
}
\label{fig:K2transit}
\end{center}
\end{figure}

Due to the sparse time sampling of the
K2 data ($\sim 0.5$ hour) in comparison with the transit duration
($\sim 1$ hour), the folded K2 transit curve looks V-shaped as shown
in Figure \ref{fig:K2transit}. This leads to a degeneracy in system
parameters (i.e., the scaled semi-major axis $a/R_s$, transit impact
parameter $b$, and planet-to-star radius ratio $R_p/R_s$) when we fit
the time-integrated K2 flux data alone.  On the other hand, the follow-up
transit curves exhibit a clearer ingress (egress) and flat bottom
(especially in the $z_\mathrm{s,2}$-band) in spite of the worse
photometric precision than {\it Kepler}, suggesting that the
transiting body is significantly smaller than the central late-type
star. The clear ingress and egress also enable us to better
constrain the transit parameters, 
and thus we here decide to fit the follow-up transit curves simultaneously 
with the K2 light curve.

When a source of dilution, i.e. a physical companion star or
background star, is present in the photometric aperture, the transit
depths measured in different bands may vary depending on the contrast
ratio of the objects in each band. Conversely, a
wavelength-independent transit depth is suggestive of no dilution
source in the aperture, on the assumption that the diluting star does
not have an identical spectral energy distribution as that of the
primary. Thus, in order to constrain the presence of possible dilution
sources, we attempt to measure the radius ratio for each observed band
as accurately as possible, and compare the results in different bands
from the optical to the near infrared. 
Since only a part of the transit is observed for OAO datasets, 
we do not attempt to fit the light curves in individual bands, but 
combine all the transit curves and perform a global fit.

We employ the method of Gaussian processes (GP) to obtain the most
accurate radius ratio for each band from our current datasets. In
addition to the flux counts, ground-based observations in general
provide many auxiliary variables such as target's pixel centroid
drifts, sky background count, target's FWHM in the photometric aperture, etc., 
which are not monotonic functions of time. Time-correlated noises are also present in our datasets
owing both to the intrinsic stellar activity and other instrumental
systematics. Taking account of these pieces of information in modeling
the observed light curves yields more accurate estimates
for the model parameters \citep[e.g.,][]{2012MNRAS.419.2683G}.  A GP
model assumes that the observed flux values with $N_\mathrm{data}$ points
follow a multi-variable Gaussian:
\begin{eqnarray} \label{eq:GP} 
\mathcal{N}(\mathbf{\mu},\Sigma)=\frac{1}{\sqrt{(2\pi)^{N_\mathrm{data}}|\Sigma|}}\exp
\left\{
-\frac{(\mathbf{f}-\mathbf{\mu})^\mathrm{T} 
\Sigma^{-1}
(\mathbf{f}-\mathbf{\mu})}{2}
\right\},~~
\end{eqnarray} 
where $\mathbf{f}$ and $\mathbf{\mu}$ are the observed and
modeled flux vectors (with $N_\mathrm{data}$ components), and $\Sigma$ is the covariance matrix.
When $\Sigma$ has only diagonal components, representing independent Gaussian noises
in individual fluxes, the exponent in Equation (\ref{eq:GP}) reduces to $-\chi^2/2$. 
By introducing non-diagonal components in the covariance matrix $\Sigma$, 
we can deal with correlated noises among flux values, not only as a function of 
time but also functions of other auxiliary parameters like pixel centroid shifts and sky background
counts, which are often corrected by modeling the baseline flux 
variations with e.g., polynomials of the parameters. In a GP modeling, we do not need to 
assume such a functional form for the flux variation by auxiliary variables, 
and the best correlation pattern among the flux values are found in the fitting process
through an optimization of hyper-parameters \citep{Rasmussen}. 

We estimate the posterior distribution of the system parameter vector $\alpha$ 
using Bayes' theorem:
\begin{eqnarray}
\label{eq:bayes}
p(\mathbf{\alpha} |\mathrm{data}) \propto \mathcal{L}(\mathrm{data}| \mathbf{\alpha})
\cdot p(\mathbf{\alpha}),
\end{eqnarray}
where $\mathcal{L}(\mathrm{data}| \mathbf{\alpha})$ is the likelihood of the flux values 
and $p(\mathbf{\alpha})$ is a prior distribution for the system parameters. 
In the present case, where we simultaneously model the observed fluxes in five
different bands with GP, the likelihood $\mathcal{L}$ is expressed as a product of 
Gaussians given in Equation (\ref{eq:GP}) for individual bands: 
\begin{eqnarray} \label{eq:likelihood} 
\ln
\mathcal{L}(\mathrm{data}| \mathbf{\alpha}, \mathbf{\theta})
&=&-\frac{1}{2}
\sum_{i=r'_2,z_\mathrm{s,2},J,H,K_\mathrm{s}}\{
(\Delta
\mathbf{f}^{(i)})^\mathrm{T} (\Sigma^{(i)}(\mathbf{\theta}))^{-1}
\Delta \mathbf{f}^{(i)}
\nonumber \\ 
&& + \ln |\Sigma^{(i)}(\theta))| +
N_{\mathrm{data}}^{(i)}\ln (2\pi)\} 
\nonumber \\ 
&& -\frac{1}{2}\chi_\mathrm{K2}^2 + \mathrm{const.}, 
\end{eqnarray} 
where $\Delta \mathbf{f}^{(i)} = \mathbf{f}^{(i)} - \mathbf{\mu}^{(i)}(\mathbf{\alpha})$, and $\mathbf{f}^{(i)}$ and
$\mathbf{\mu}^{(i)}$ are $i-$th band's observed and model flux vectors
each comprised of $N_{\mathrm{data}}^{(i)}$ rows, and
$\Sigma^{(i)}(\mathbf{\theta})$ is the covariance matrix 
of $N_{\mathrm{data}}^{(i)}\times N_{\mathrm{data}}^{(i)}$ for that dataset. 
The vectors $\mathbf{\alpha}$ and $\mathbf{\theta}=\{A_{j}^{(i)}, L_{j}^{(i)}\}$
are transit model parameters and hyper-parameters, respectively, which
are to be optimized by the procedure below.  For the model flux
$\mathbf{\mu}^{(i)}$, we use the analytic model by \citet{2009ApJ...690....1O}.
Since we have already corrected for the pixel centroid motion and time-dependent flux variation
in extracting the reduced K2 light curve, 
we employ a simple $\chi^2$ statistics for the K2 dataset as
\begin{eqnarray}
\label{eq:chisq}
\chi_\mathrm{K2}^2 = \sum_{i}\frac{(f_\mathrm{LC,obs}^{(i)}-f_\mathrm{LC, model}^{(i)})^2}{\sigma_\mathrm{LC}^{(i)2}},
\end{eqnarray}
where $f_\mathrm{LC,obs}^{(i)}$ and $\sigma_\mathrm{LC}^{(i)}$ are the $i$-th observed K2 flux and its error, respectively. 
For $f_\mathrm{LC,obs}^{(i)}$, 
we extracted light curve segments only around the transit 
(covering $\pm 2$ times the transit duration from the mid-transit time) from the full reduced K2 light curve 
to save the computation time. 
Note that $f_\mathrm{LC, model}^{(i)}$ is computed by integrating the transit model flux \citep{2009ApJ...690....1O} 
over the cadence of K2 observation ($\sim 29.4$ minutes).

For the covariance matrix $\Sigma$, there are some choices to describe the correlations between 
flux values and auxiliary parameters. 
Here we simply adopt the following combination of white noises and
the ``squared exponential" kernels:
\begin{eqnarray}
\label{eq:covariant}
\Sigma^{(i)}_{nm} = \sum_{j=1}^5A_{j}^{(i)2}\exp\left(-\frac{(p_{j,n}^{(i)}-p_{j,m}^{(i)})^2}{2L_{j}^{(i)2}}\right)
+\delta_{nm}\sigma_{n}^{(i)2},
\end{eqnarray}
where the first term corresponds to the correlation between the input variables (parameters) $p_{j,n}^{(i)}$ and $p_{j,m}^{(i)}$,
while the second term is the white noise component in each of the observed fluxes. 
As auxiliary variables $p_{j,n}^{(i)}$, we here employ the time $t$, $x$ and $y$ pixel centroid drifts (each), 
sky background count, and target's FWHM in the photometric aperture, 
in total introducing ten hyper-parameters for each bandpass. 
In this specific covariance matrix, 
we can ``learn" from the data the amplitude and length (scale) of the flux correlation as a function of each auxiliary variables
by optimizing the hyper-parameters $A_j^{(i)}$ and $L_j^{(i)}$. 
We do not incorporate the target's airmass as an auxiliary parameter, since target's airmass varied monotonically 
against time during both IRSF and OAO runs, implying that GP regressions by target's airmass could be
degenerate with those by time (red noise). As we show in Section \ref{s:discussion}, however, fitting the light curves including 
airmass terms in the GP regression yields a fully consistent result with the one without airmass terms.

\begin{table*}[tb]
\begin{center}
\caption{Result of the Global Fit to Transit Light Curves}\label{hyo2}
\begin{tabular}{lcccccc}
\hline
Parameter & $r'_2$-band & $z_\mathrm{s,2}$-band & $J$-band & $H$-band 
& $K_\mathrm{s}$-band& Kepler-band\\\hline\hline
\multicolumn{7}{l}{\it (Fitting Parameters in individual bands)} \\
$u_1+u_2$ & $0.78\pm 0.10$ & $0.64\pm 0.10$ & $0.45\pm 0.10$ & $0.44\pm0.10$& $0.34\pm 0.10$ & $0.080\pm 0.09$\\
$u_1-u_2$ & $0.13\pm 0.10$ & $0.43\pm 0.10$ & $-0.29\pm 0.10$ & $-0.32\pm0.10$& $-0.28\pm 0.10$ & $0.00\pm 0.10$\\
$R_p/R_s$ & $0.056_{-0.010}^{+0.009}$ & $0.077_{-0.004}^{+0.005}$ & $0.063\pm 0.007$ & $0.073\pm 0.007$ & $0.086_{-0.006}^{+0.005}$ & $0.0737_{-0.0018}^{+0.0032}$\\
\hline
\multicolumn{7}{l}{\it (Common Fitting Parameters)} \\
$a/R_s$ & \multicolumn{6}{c}{$17.9_{-2.8}^{+1.2}$} \\
$b$ & \multicolumn{6}{c}{$0.36_{-0.24}^{+0.26}$} \\
$e$ & \multicolumn{6}{c}{$0$ (fixed)}\\
$P$ (days)& \multicolumn{6}{c}{$2.260455\pm 0.000041$} \\
$T_c^{\mathrm{(IRSF)}}$ &&& \multicolumn{3}{c}{$2457242.4620_{-0.0048}^{+0.0049}$} &\\
$T_c^{\mathrm{(OAO)}}$ & \multicolumn{2}{c}{$2457258.2849_{-0.0051}^{+0.0052}$} &&&&\\
$T_c^{\mathrm{(K2)}}$ &&&&&& $2456977.99012_{-0.00074}^{+0.00075}$ \\
\hline
\end{tabular}
\end{center}
\end{table*}
In the global fit to the light curves, we have the following system model parameters:
$a/R_s$, $b$, $P$, the limb-darkening parameters $u_1$ and $u_2$ for the quadratic limb-darkening law, 
$R_p/R_s$, and times of mid-transit for OAO, IRSF, and K2 datasets ($T_c^{\mathrm{(OAO)}}$, $T_c^{\mathrm{(IRSF)}}$,
and $T_c^{\mathrm{(K2)}}$) as summarized in Table \ref{hyo2}. 
Among these, $a/R_s$, $b$, $P$, $T_c^{\mathrm{(OAO)}}$, $T_c^{\mathrm{(IRSF)}}$, $T_c^{\mathrm{(K2)}}$ 
are common to all datasets, 
but we allow the other parameters to vary to see the possible variation in $R_p/R_s$ for each band. 
Due to the quality of the data, we are forced to fix the orbital eccentricity at zero, and also impose
Gaussian priors (with dispersions of $0.1$ for both $u_1+u_2$ and $u_1-u_2$) on the limb-darkening parameters 
based on the table provided by \citet{2013A&A...552A..16C} as $p(\mathbf{\alpha})$ in Equation (\ref{eq:bayes}). 
To take into account the case that the internally estimated white noise for each flux value 
($\sigma_{n,\mathrm{internal}}^{(i)}$: photon plus scintillation noise) is underestimated, 
we also optimize the white noise component $\sigma_{n}^{(i)}$ in Equation (\ref{eq:covariant})
by introducing additional free parameters
$\sigma_{\mathrm{white}}^{(i)}$ for individual bands via 
\begin{eqnarray}
\label{eq:white}
\sigma_{n}^{(i)} = \sqrt{\sigma_{n,\mathrm{internal}}^{(i)2} + \sigma_{\mathrm{white}}^{(i)2}}. 
\end{eqnarray}

On the basis of Bayesian framework, we estimate the marginalized posteriors for those parameters. 
Ideally, the posterior distributions for these fitting parameters should be inferred by marginalizing
all of the system and hyper-parameters. But the size of data and the huge number of parameters
prohibit the full marginalization: computation of the inverse covariance matrix is rather expensive. 
Therefore, following \citet{2015MNRAS.451..680E}, we decide to adopt the so-called type-II maximum likelihood as below. 
We first maximize Equation (\ref{eq:likelihood}) by the Nelder-Mead simplex method \citep[e.g.,][]{2002nrc..book.....P}, varying 
all the model and hyper-parameters. We then fix the hyper-parameters and $\sigma_{\mathrm{white}}^{(i)}$ in 
Equation (\ref{eq:white}) at the optimized values, and 
run Markov Chain Monte Carlo (MCMC) simulations using the customized code 
\citep{2012ApJ...759L..36H, 2015ApJ...799....9H}
to obtain the global posterior distribution. 
The step size for each parameter is iteratively optimized so that the global acceptance ratio 
falls between $10-40~\%$. We run $10^7$ MCMC steps and the representative values are 
extracted from the marginalized posterior for each system parameter by taking the median, and 15.87 
and 84.13 percentiles as the best-fit value and its $\pm 1\sigma$. 
We list the result of the fit in Table \ref{hyo2}. 
The best-fit light curve models after subtracting the GP regressions to the correlated noises
are displayed in the bottom panels of Figure \ref{fig:irsf_jhk} and \ref{fig:oao_rz} 
for the IRSF and OAO datasets, respectively. 
We note that the optimized $\sigma_{\mathrm{white}}$ is typically $0.001-0.003$.



\begin{table}[tb]
\begin{center}
\caption{Final Planetary Parameters}\label{hyo3}
\begin{tabular}{lc}
\hline
Parameter & Value \\\hline\hline
$P$ (days)& $2.260455\pm 0.000041$ \\
$T_{c,0}$ (BJD) & $2456977.99012_{-0.00074}^{+0.00075}$\\
$R_p$ ($R_\oplus$) & $2.32\pm 0.24$ \\
$i_o$ ($^\circ$) & $88.9_{-1.2}^{+0.8}$ \\
$a$ (AU) & $0.0214\pm0.0013$ \\
$T_\mathrm{eq}$ (K) (Bond albedo: 0.0)& $568\pm 35$ \\
$T_\mathrm{eq}$ (K) (Bond albedo: 0.4)& $500\pm 31$ \\
\hline
\end{tabular}
\end{center}
\end{table}
\begin{figure}[t]
\begin{center}
\includegraphics[width=8.5cm]{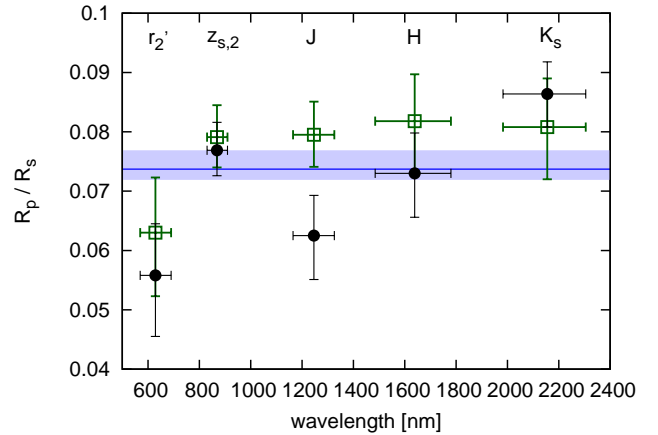} 
\caption{Planet-to-star radius ratio ($R_p/R_s$) in each observed band (filled circles). 
The blue area represents $\pm 1\sigma$ of $R_p/R_s$ in the Kepler band. 
The horizontal errorbar for each band corresponds to the range of wavelength
covered by the band. Note that the Kepler band covers the optical region between 
$437-835$ nm. 
The open squares are the measured $R_p/R_s$ for the cases that 
airmass cutoffs ($<2.2$) are applied (the $r_2^\prime$ and $z_{s,2}$ bands)
and light curves are extracted with time-variable apertures (the $J$, $H$, and $K_s$ bands) in the fit. 
}
\label{fig:rprs}
\end{center}
\end{figure}
Figure \ref{fig:K2transit} shows the phase-folded K2 data (black points) along with our best-fit model (red solid line),
and Table \ref{hyo3} summarizes our final result for the system parameters. 
Comparing the radius ratio by K2 data analysis with those by ground-based observations,
$R_p/R_s$ from the optical to the infrared is consistent within $\sim 2\sigma$ 
(Figure \ref{fig:rprs}: filled circles). 
The good agreement between the K2 transit depth and that in the $z_\mathrm{s,2}$ band,
in which the best photometric precision was achieved from the ground, 
suggests that the transit-like signal is not caused by a background/bound eclipsing binary. 
Nonetheless, the transit depths in the $r_2^\prime$- and $K_\mathrm{s}$-bands exhibit a moderate
disagreement. We will revisit this issue in Section \ref{s:discussion}.

The relatively short transit duration of \target b, in spite of the moderate impact parameter,
suggests that the stellar radius is small and thus the star has a higher density. 
Indeed, using $a/R_s=17.9_{-2.8}^{+1.2}$ and Kepler's third law, we estimate the 
stellar density as $\rho_\star/\rho_\odot=15.1_{-5.9}^{+3.2}$ solely from the transit light curve. 
Comparing this value with the spectroscopic estimate (Table \ref{hyo1}), we find that
they are compatible with each other, making it highly likely that 
\target b is transiting a cool star.

\section{Validation of the Candidate\label{s:astrovalid}}\label{s:astrovalid}

\subsection{Resolved Sources in the Field \label{s:astrometry}}\label{s:astrometry}

\begin{figure}[t]
\begin{center}
\includegraphics[width=7.0cm]{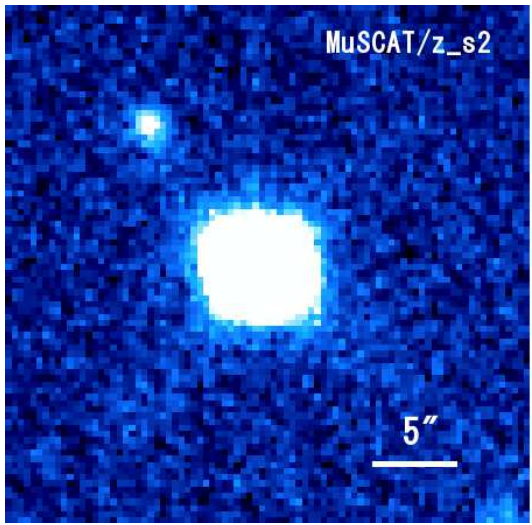} 
\caption{MuSCAT's on-focus image for \target\ in the $z_\mathrm{s,2}$ band. 
North is up and east is left. Linear scaling is applied in this image. 
}
\label{fig:muscat_onfocus}
\end{center}
\end{figure}
As we noted, the SDSS image taken in 2000 suggests that \target\ has a faint companion
to its north-east at a separation of $\sim 5\farcs2$, which could become a source of
false positive. 
Figure \ref{fig:muscat_onfocus} plots our latest $z_\mathrm{s,2}$-band image taken by MuSCAT on 2015 August 23 UT, 
in which we found the faint companion 
with the same magnitude difference as in SDSS, located further from \target. 
Considering the proper motion of \target\ ($\mu_\alpha=-256\pm3$ mas yr$^{-1}$ 
and $\mu_\delta=-195\pm 3$ mas yr$^{-1}$) \citep{2012ApJS..203...21A}, the expected current separation between 
the two is $\sim10^{\prime\prime}$ on the assumption that the fainter object is a background one. 
After calibrating the coordinates of both stars, we find that the faint companion stays at 
the almost same location while the current coordinate of \target\ has slightly moved to
$\mathrm{RA}=22:22:29.59,~\mathrm{Dec}=-07:57:22.78$. 
The current separation between the two is estimated as $10\farcs36 \pm 0\farcs17$, 
which is fully consistent with the background scenario of the faint object. 

Although the coordinate of the background star is outside of our custom-made aperture
in the K2 photometry, a portion of its pixel response function (PRF) is involved in the aperture, 
meaning that the background star could still be a source of false positive. 
We checked the magnitude of its contamination on \target\ assuming the averaged PRF
for {\it Kepler} prime mission \citep[Equation (10) in][]{2014AJ....147..119C}. To estimate the fraction of the companion's PRF
that falls on our custom-made aperture, we simplified the aperture so that it is a circle encompassing all the
aperture pixels with its radius being $8\farcs44$. The resulting companion's PRF fraction was estimated as
$\sim 0.248$, and combining this with the magnitude difference between the two stars 
($\Delta m_\mathrm{kep}\approx4.82$), we estimated the maximum contamination from 
this companion as $\approx 3.0\times10^{-3}$. 
The PRF for K2 could be larger than for the {\it Kepler} prime mission, but
since the aperture does not extend to the coordinate of the background star, the PRF
fraction cannot exceed $\sim 0.5$. 
Along with the fact that the maximum flux contamination from the companion star is smaller than 
the observed transit depth ($0.006-0.007$), 
we conclude that it is not responsible for the transit-like signal detected in our pipeline. 
Note that we also checked the current coordinate of \target\ in the SDSS image, finding no bright object
which could become a possible source of false positives. 

\subsection{Bayesian Statistical Calculation \label{s:validation}}\label{s:validation}

We performed a Bayesian calculation of the false positive probability (FPP) that the signal arises from a background star (i.e., an eclipsing binary, EB) in the vicinity of the location of \target.  The calculation does not address the probability that such a star is actually a binary on an eclipsing orbit, 
but address the probability that
an appropriate star is close on the sky to produce the signal, and thus it is an upper limit on FPP. The procedure is described in detail in Gaidos et al. submitted, ApJ, and only briefly described here.  The calculation multiplies a prior probability based on a model of the background stellar population by likelihoods from observational constraints.  The synthetic background population at the location of \target\
was constructed using TRILEGAL Version 1.6 \citep{Vanhollebeke2009}: to improve counting statistics, the population equivalent to 10 sq. deg. was computed.  The background was computed to $m_\mathrm{kep} = 23$, i.e. far fainter than the faintest object ($m_\mathrm{kep}\approx 19.5$) that could produce the signal if it were an EB with the maximum eclipse depth of 50\%.  The likelihood factors are the probabilities that (a) the background star can produce the observed transit depth; (b) the mean density of the background star is consistent with the observed transit duration; and (c) the background star does not appear in our Subaru ICRS-AO $K^\prime$-band imaging of the \target\ (Section \ref{s:ircs}).  

The calculation was performed by Monte Carlo: it sampled the synthetic background population randomly and placed them randomly and uniformly over a $15^{\prime\prime}$-radius circle centered on \target.  
Stars that violated the AO contrast ratio constraint (condition c) were excluded.  
Given the known orbital period and mean density of the synthetic star, the probability that a binary would have an orbit capable of producing the observed transit duration (condition b) was calculated assuming a Rayleigh distribution of orbital eccentricities with mean of 0.1.  (Binaries on short-period orbits should quickly circularize.)\footnote{The eclipse duration calculation uses the formula for a ``small'' occulting object and so is only approximate.}  To determine whether a background star could produce the observed transit signal with an eclipse depth $<50\%$ (condition a), we determined the relative contribution to the flux of \target\ using bilinear interpolations of the PRF for detector channel 48 with the tables provided in the Supplement to the Kepler Instrument Handbook (E. Van Cleve \& D. A. Caldwell, KSCI-19033).  The calculations were performed in a series of 1000 Monte Carlo iterations and a running average used to monitor convergence.  
We found a FPP of $\approx 2 \times 10^{-5}$ and therefore, we rule out the false positives by a background eclipsing binary.

\subsection{Constraint on Possible Dilution Sources \label{s:validation2}}\label{s:validation2}

The remaining false positive case is that a physically associated stellar companion is present around \target. 
The bound companion could have a transiting object (case A), or \target\ could have a transiting object but its depth is diluted by 
the bound companion (case B). Our AO image achieves a $5\sigma$ contrast of $\sim 0.24$ at a separation of $0\farcs1$, which
translates to $\sim 5$ AU from \target. Thus, it is possible that a bound companion later than M4 dwarf is
present within this distance from the central star. However,
the similar values for the transit depth ($R_p/R_s$) from the optical to near infrared suggest that case A
is unlikely, when the bound companion has a different spectral type from \target.

To quantify this statement, we refit the observed light curves introducing a ``dilution factor" $D$,
defined as the ratio of the companion's flux to that of \target\ in each bandpass. 
The companion has to be equal to or later than M4 dwarf since our spectroscopy implies \target\ is the
dominant source of brightness in the system. 
Thus, $D$ is in general larger in the infrared than in the optical. 
On the assumption that the later-type companion has a transiting object (case A), 
we search for a solution to the observed light curves as in Section \ref{s:analysis}. 
We assume various stellar types for the companion, and employ the contrast ratio for each band from \citet{2007AJ....134.2340K}. 
We simply use the $r$ and $z$ magnitudes in \citet{2007AJ....134.2340K} to represent the MuSCAT $r'_2$ and $z_\mathrm{s,2}$ 
bands, and $m_\mathrm{kep}$ is computed by $0.1m_g+0.9m_r$. 
As a result of the global fit to the ground-based transit follow-ups along with 
the K2 light curve including $D$, 
we find that a companion later than M4 dwarf leads to an incompatible result: 
in case of an M5 dwarf companion, 
the intrinsic radius ratio of the eclipsing objects in the optical 
(e.g., $R_p/R_s=0.206\pm 0.003$ for the Kepler band) becomes inconsistent with that  
in the infrared (e.g., $R_p/R_s=0.134\pm 0.008$ for the $J$ band) with $>5\sigma$. 
Hence, the putative bound companion (having a transiting object) around \target, if any, 
has to be another M4 dwarf. Even in this case, the transiting object falls on the planetary regime
considering that the maximum possible dilution (M4+M4 binary case) brings about an underestimation 
of the radius ratio by a factor of $\sim \sqrt{2}$.

For the rest of the discussion, we resort to the statistics to constrain the possible dilution scenario
by computing the probability that ``\target\ has an almost identical bound stellar companion". 
First, the probability that an M dwarf ($0.1-0.5M_\odot$)
has any stellar companion is $26\pm3~\%$, and the probability that their mass ratio $q$ is
greater than 0.75 ($\approx$ mass ratio between M4 and M5 dwarfs) is $0.57_{-0.27}^{+0.17}$ 
on the assumption that the probability distribution of $q$ follows $\propto q^\gamma$ with
$\gamma=1.9\pm1.7$ \citep{2013ARA&A..51..269D}. Then, adopting the log-normal distribution for the period $P$, 
we estimate the probability that binary's semi-major axis is smaller than 5 AU as $\approx0.49$ \citep{2013ARA&A..51..269D}. 
Thus, the total probability that \target\ has such an M4 companion is $\approx 0.0721_{-0.036}^{+0.023}$. 
This is not critically low, yet we can safely say that it is more likely that
\target\ is a single star with a transiting super-Earth/mini-Neptune\footnote{
The division between super-Earths and mini-Neptunes is still ambiguous. 
Planets with $M_p<10M_\oplus$ are conventionally referred to as super-Earths
and \target's radius ($\approx 2.32R_\oplus$) suggests that its mass is smaller than
$10M_\oplus$. But the division could also depend on the host star's type and 
orbital period.}.

We note that case B of the dilution scenario is also possible, but this possibility is 
not so high as well following the same discussion above. In order to estimate the maximum 
radius ratio, we repeat the fit of the light curves with $D=1.0$, representing 
the case that a bound M4 dwarf identical to \target\ is present in the system.
The global fit to the observed light curves yields $R_p/R_s=0.104_{-0.002}^{+0.004}$, 
corresponding to $R_p\approx3.27R_\oplus$. 
Again, this is an upper limit of the planet radius, and the planetary size is much closer
to $R_p\approx 2.32R_\oplus$ when the dilution source is later than M4. 
All these dilution possibilities could be settled by taking a high resolution spectrum of the target
and checking the binarity from the line blending, although the faintness of the target 
would make it challenging in the optical region ($m_V\approx16.1$).

\section{Discussion and Summary \label{s:discussion}}\label{s:discussion}
We have conducted intensive follow-up observations for \target, which emerged as a
planet-host candidate within the ESPRINT collaboration. 
Our optical spectroscopy indicates that \target\ is a metal-rich M4 dwarf, located at
$53\pm8$ pc away from us. Based on the absence of bright sources 
in the AO image taken by Subaru/IRCS, we computed the probability that the transit-like signal
is caused by a background eclipsing binary, and showed that such an FPP is very low
($\approx 2 \times 10^{-5}$). The remaining possible false positive scenario is that a physically 
associated companion has a transiting object, but this still puts \target b in the planetary regime 
considering the maximum possible dilution case. 
Our ground-based transit follow-ups using OAO/MuSCAT
and IRSF/SIRIUS revealed similar transit depths in different bands from the optical to the near infrared, 
thus showing such a bound companion is not likely to exist around \target, with the probability
being $\approx 0.0721_{-0.036}^{+0.023}$. 

It should be emphasized that the high cadence photometry of ground-based follow-ups helped
to break the degeneracy between the system parameters. The poor sampling of the K2 data makes 
the transit curve V-shaped as shown in Figure \ref{fig:K2transit}, which could be explained by
a grazing eclipsing binary, but follow-up transits exhibit flat bottoms
in all bands, which are suggestive of a small transiting body. Even in the absence of a prior probability on the 
stellar density in fitting the follow-up transits, we could obtain a relatively tight constraint on the
scaled semi-major axis and radius ratio. This case clearly demonstrates the importance of transit follow-ups
from the ground to validate planetary candidates with relatively short transit durations.

Despite the careful analysis for the follow-up transits using GP, the transit depths in the $r_2^\prime$ and 
$K_\mathrm{s}$ bands shows a moderate disagreement ($\sim 2\sigma$) with that in the 
Kepler band as shown in Figure \ref{fig:rprs}. 
As discussed below, these differences in $R_p/R_s$ are significantly larger than the ones expected from the 
different optically-thick planet radii in individual bands ($\Delta(R_p/R_s)\approx 0.003$ at the most
assuming a hydrogen-rich atmosphere). 
The disagreement in the $r_2^\prime$ band could be due to the lack of egress combined
with the small number of data points during the transit. 
At the end of the OAO/MuSCAT run, the target was also low in elevation, suggesting that
higher airmass may have caused some systematics. To take account of the airmass-related
systematics, we also performed a global fit to the observed light curves including an airmass-dependent 
GP term in Equation (\ref{eq:covariant}). The result of the fit was fully consistent with the result without the airmass-dependent
GP term (i.e., $R_p/R_s=0.057_{-0.010}^{+0.009}$ for the $r_2^\prime$ band). 
On the other hand, if we simply remove the flux data taken at higher airmass ($>2.2$) 
from the OAO dataset,
the global fit yields $R_p/R_s=0.063_{-0.011}^{+0.009}$ for the $r_2^\prime$ band, which agrees with 
$R_p/R_s$ in the Kepler band with $\sim 1\sigma$. 
This treatment is rather arbitrary so that we do not claim its result as the final one. 
Further transit observations covering a whole transit would be able to settle this issue.

We also investigated the reason for the disagreement in the $K_\mathrm{s}$ band by checking the raw images 
taken by IRSF. We found that the FWHM of the target image slightly changed 
(by $\sim 20\%$) during our IRSF observation, and that variation was not a monotonic function of time;
the FWHM takes its maximum during the transit. 
Although we have incorporated the FWHM as an input auxiliary variable in the GP regression
to the correlated noises, this significant variation in FWHM may have caused further systematics 
in the extracted light curves. 

To further investigate this possibility, we adopt time-variable apertures and set each aperture radius 
as FWHM multiplied by a constant value (e.g., 0.6), and extracted light curves again. 
Then, following the fitting procedure described in Section \ref{s:analysis}, we estimate the planet-to-star
radius ratio for each band. Consequently, we find 
$R_p/R_s=0.081_{-0.009}^{+0.008}$ for the $K_\mathrm{s}$ bands
(Figure \ref{fig:rprs}: open squares).
This value is consistent with the transit depth in the Kepler band ($R_p/R_s=0.074_{-0.002}^{+0.003}$), 
but the photometric precision turns out to be much worse 
than the fixed-aperture photometry. 
The reason for this discrepancy between the light curves for
fixed and time-variable apertures is not known, but imperfect correction for flat-fielding or
inclusion of scattered light from a neighboring star could be relevant.

\begin{figure}[t]
\begin{center}
\includegraphics[width=8.5cm]{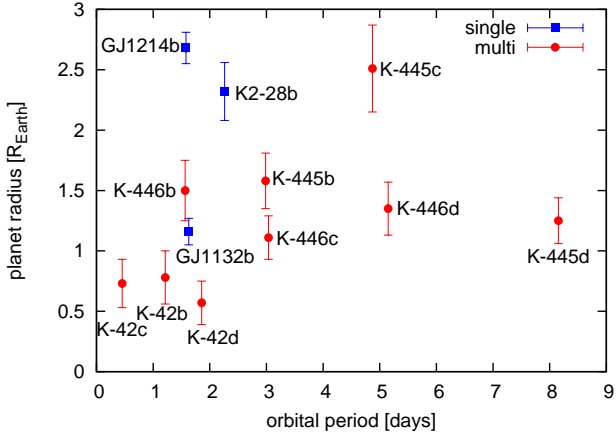} 
\caption{Transiting planets around M dwarfs cooler than $T_\mathrm{eff}\leq 3400$ K. 
Planets in single and multiple systems are shown in blue squares and red circles, respectively
\citep{2009Natur.462..891C, 2012ApJ...747..144M, 2015ApJ...801...18M, 2015Natur.527..204B}. 
``Kepler-" is abbreviated as ``K-" in this plot. 
}
\label{fig:midM_planets}
\end{center}
\end{figure}
\citet{2013ApJ...767...95D} found that while the occurrence rate of Earth-sized planets
($R_p<1.4R_\oplus$) around coolest M dwarfs monitored by {\it Kepler} is consistent with that 
around hotter M dwarfs ($T_\mathrm{eff}\geq 3723$ K), the occurrence rate of super-Earths 
around cooler M dwarfs is significantly smaller than that around hotter M dwarfs. 
Figure \ref{fig:midM_planets} plots the transiting planets around M dwarfs later than M3. 
Among those planets, the transiting planets with $R_p\geq2.0R_\oplus$ are only GJ 1214b, Kepler-445c, and \target b. 
We note that their hosts are all metal-rich stars, which is in marked contrast to
the metal poor hosts, Kepler-42, Kepler-446, and GJ 1132, having only Earth-sized (or sub-Earth-sized) planets. 
The occurrence rate calculated by \citet{2013ApJ...767...95D} could contain $T_\mathrm{eff}$-dependent
systematic errors in the stellar (and thus planet) radii, arising from the adopted models which become
unreliable for lower $T_\mathrm{eff}$. 
With more samples as presented here, one can discuss the statistical property of
planets around cooler M dwarfs more accurately.

With the relatively bright host star in the near infrared and moderate transit depth,
\target\ is a good target for future follow-up studies.  
It is particularly tempting to compare \target b with another well-studied super-Earth/mini-Neptune, GJ 1214b, 
in terms of atmospheric characteriztions. 
Using the semi-major axis of $a=0.0214\pm0.0013$ AU and \target's radius and effective temperature in Table \ref{hyo1}, 
we find that \target b happens to receive almost an equivalent insolation from its host star as GJ 1214b
(incident energy fluxes $S_\mathrm{inc}\approx17.3S_\oplus$ for \target b
and $S_\mathrm{inc}\approx17.6S_\oplus$ for GJ 1214b, respectively). 
\citet{2014Natur.505...69K} conducted a spectro-photometry with 
the Wide Field Camera 3 (WFC3) on the Hubble Space Telescope (HST),
and as a result of analyzing 12 transits of GJ 1214b, they achieved a 
precision of $30-40$ ppm for the transit depth in the individual spectroscopic channel
between $1.15-1.65\mu$m. 
Since the transit depth variation against wavelength scales
as $\Delta(R_p^2/R_s^2) \propto 2(R_p/R_s)^2(H/R_p)$, where $H$ is the scale height, 
this variation for \target\ as a function of wavelength would be 
$0.30\times$ that for GJ 1214b. 
If one conducts a similar observation for \target\
to the one conducted by \citet{2014Natur.505...69K}, observing the same 
number of transits with HST, we expect a precision of $70-80$ ppm for the transit depth measurement
on the assumption that the uncertainty is dominated by the photon-limited
shot noise ($\Delta m_H=1.94$ mag and transit duration of $\sim 60$ minutes). 
This level of precision is sufficient to confirm or rule out the atmospheres dominated
by hydrogen or methane, for which the scale height equal to GJ 1214b would lead to
$\Delta(R_p^2/R_s^2)\sim 450$ ppm and $\sim 150$ ppm in variation amplitudes, respectively. 
Ruling out the water- or carbon-dioxide-dominated atmospheres 
($\sim 80$ ppm and $\sim 20$ ppm, respectively)
could be challenging, but increasing the number of observed transits will help.

Searching for additional planets in the system is also important to understand the
architecture of planetary systems around mid-M dwarfs. 
\citet{2015ApJ...801...18M} showed that a significant fraction ($21_{-5}^{+7}~\%$)
of mid-M dwarfs hosts multiple planets within 10 days. 
Concerning \target, we could not find evidence of another transiting planet
in our BLS analysis. The sparse sampling of the K2 data during the transit
makes it complicated to search for possible transit timing variations (TTVs), but further 
intensive ground-based transit follow-ups would find or at least be able to put a constraint 
on the presence of additional bodies. 
While \target\ is faint in the optical, it is relatively bright in the NIR ($m_H = 11.03\pm 0.02$ mag),
and thus is likely within the reach of existing and planned NIR radial velocity instruments \citep[e.g., IRD, CARMENES, SPIrou, HPF;][]{2014SPIE.9147E..14K, 2014SPIE.9147E..1FQ, 2014SPIE.9147E..15A, 2014SPIE.9147E..1GM}, 
which could also reveal any non-transiting planets.

\acknowledgments 

This paper is based on data collected at Subaru Telescope, which is
operated by the National Astronomical Observatory of Japan.  
We are grateful to Josh Winn and Simon Albrecht for the discussions
within the ESPRINT collaboration. 
We acknowledge Kumiko Morihana and Qazuya Wada for a support of our IRSF observations.
We thank the Subaru support astronomers, Dr. Tae-Soo Pyo and Dr. Joanna Bulger, 
for their help to carry out the observations.
The data analysis was in part carried out on common use data analysis computer system 
at the Astronomy Data Center, ADC, of the National Astronomical Observatory of Japan. 
T.H.\ and M.K.\ are supported by Japan Society for Promotion of Science (JSPS) 
Fellowship for Research (No. 25-3183 and 25-8826). 
A.F.\ acknowledges support by the Astrobiology Center Project of National Institutes of 
Natural Sciences (NINS) (Grant Number AB271009).
This work was performed, in part, under contract with the Jet Propulsion Laboratory (JPL) funded by 
NASA through the Sagan Fellowship Program executed by the NASA Exoplanet Science Institute. 
N.N.\ acknowledges support by the NAOJ Fellowship, Inoue Science Research Award, 
and Grant-in-Aid for Scientific Research (A) (No. 25247026) 
from the Ministry of Education, Culture, Sports, Science and Technology (MEXT) of Japan.
H.O. and M.T.\ acknowledge support by
Grant-in-Aid for Scientific Research (No.15H02063).
I.R. acknowledges support from the Spanish Ministry of Economy and Competitiveness (MINECO) 
and the Fondo Europeo de Desarrollo Regional (FEDER) through grants ESP2013-48391-C4-1-R and ESP2014-57495-C2-2-R.
We acknowledge the very significant cultural role and reverence that the
summit of Mauna Kea has always had within the indigenous people in Hawai'i. 
We express special thanks to the anonymous referee for the 
helpful comments and suggestions on this manuscript.




\end{document}